\documentclass[epj]{webofc}
\usepackage[varg]{txfonts}   
\usepackage{graphicx,color} 
\usepackage{bm}       
\usepackage{amsmath}  
\usepackage{amssymb}
\woctitle{21st International Conference on Few-Body Problems in Physics}
\begin{document}
\title{Ab initio calculation of the electromagnetic and neutral-weak response functions of $^4$He and $^{12}$C}
\author{A.\ Lovato\inst{1}\fnsep\thanks{\email{lovato@anl.gov}} \and
        O.\ Benhar \inst{2}\fnsep\thanks{\email{benhar@roma1.infn.it}} \and
        J.\ Carlson\inst{3}\and
       S.\ Gandolfi \inst{3} \and
       Steven\ C.\ Pieper\inst{1}\and
        N.\ Rocco \inst{2}\and
        R.\ Schiavilla\inst{3,4}
}

\institute{Physics Division, Argonne National Laboratory, Argonne, IL 60439
\and
           INFN and Department of Physics, ``Sapienza'' University, I-00185 Roma, Italy
\and
           Theoretical Division, Los Alamos National Laboratory, Los Alamos, NM 87545 
\and
           Theory Center, Jefferson Lab, Newport News, VA 23606
\and
           Department of Physics, Old Dominion University, Norfolk, VA 23529
          }

\abstract{Precise measurement of neutrino oscillations, and hence the determination
of their masses demands a quantitative understanding of neutrino-nucleus interactions.
To this aim, two-body meson-exchange currents have to be accounted for along within
realistic models of nuclear dynamics. We summarize our progresses towards the construction
of a consistent framework, based on quantum Monte Carlo methods and on the spectral 
function approach, that can be exploited to accurately describe neutrino interactions
with atomic nuclei over the broad kinematical region covered by neutrino experiments. 
}
\maketitle
\section{Introduction}
\label{intro}
The description of neutrino interactions with nuclei, besides being interesting in its own right, provides an essential input for neutrino-oscillation experiments,
such as DUNE \cite{LBNE:2013}, which plan to measure neutrino oscillation parameters and the neutrino mass hierarchy and determine the charge-conjugation and parity (CP) violating phase. Such experiments make use of nuclear targets as detectors that, while allowing for a substantial increase in the event rate, demand a quantitative understanding of neutrino-nucleus interactions. 

The MiniBooNE collaboration has reported a measurement of the charged-current quasielastic (CCQE) neutrino-carbon inclusive double differential cross section, exhibiting a large excess with respect to the predictions of the relativistic Fermi gas model \cite{Aguilar:2008}. It has been suggested that the discrepancy might be due to the occurrence of events with two particle-two hole final states \cite{PhysRevC.80.065501,Nieves:2011yp} that are not taken into account by the relativistic Fermi gas. Within a realistic model of nuclear dynamics, their occurrence arises naturally owing to two-body meson-exchange currents and/or correlations induced by the nuclear interaction.

\begin{figure}
\centering
\includegraphics[width=7cm,clip]{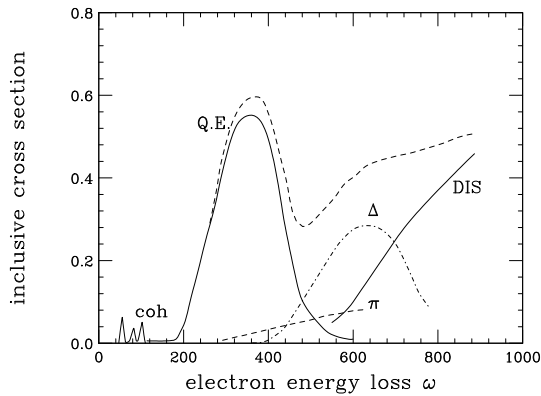}
\caption{Schematic representation of the inclusive electron-nucleus cross section at beam energy around 1 GeV, as a function
of energy loss, from Ref. \cite{Benhar:2006wy}.}
\label{fig-1}       
\end{figure}

Because neutrino beams are always produced as secondary decay products, their energy is not sharply defined, but broadly distributed. Therefore, the observed cross section for a given energy and angle of the outgoing lepton includes contributions from energy- and momentum-transfer regions where different mechanisms are at play. They can be best identified in electron-scattering experiments, in which the energy of the incoming electron is precisely known. The typical behavior of the double differential inclusive cross section for the process
\begin{equation}
e+A\to e^\prime + X\, ,
\end{equation}
with a beam energy around 1 GeV where only the outgoing electron is detected, is shown in Figure~\ref{fig-1}, taken from Ref. \cite{Benhar:2006wy}. The target nucleus in its ground state and the undetected hadronic final state are denoted by $A$ and $X$, respectively.
At small energy transfer the structure of the low-lying energy spectrum and collective effects are important. In the quasielastic peak region, the cross section is dominated by scattering off individual nucleons although a significant contribution from nucleon pairs is also present. The $\Delta$-resonance region is characterized by one or more pions in the final state, while at large energy transfer the deep inelastic scattering contribution is the largest.

The ground state of the nucleus does not depend upon the momentum transfer of the process and can be accurately described within nonrelativistic many-body theory (NMBT). On the other hand, the final state does depend on the momentum transfer and the nonrelativistic approximation can be safely applied only in the region of low and moderate momentum transfer, corresponding to $|\mathbf{q}|\lesssim 500$ MeV.  The treatment of the region of high momentum transfer requires a theoretical approach in which the accurate description of the nuclear ground state provided by NMBT is combined with a relativistically consistent description of both the final state and the nuclear current.

Setting up a framework in which two-body currents and nuclear correlations induced by realistic Hamiltonians are consistently accounted for is the main topic of this proceeding.

\section{Lepton-Nucleus cross section}
The double differential inclusive cross-section of the electromagnetic (EM) and neutral-current (NC) lepton-nucleus transition $\ell+A\to\ell^\prime +X$ is proportional to the product of the leptonic and the hadronic tensors,
\begin{equation}
\frac{d\sigma}{d\Omega_\ell d E_\ell} \propto L_{\mu\nu}W^{\mu\nu}
\end{equation}
The leptonic tensor $L_{\mu\nu}$ is completely determined by lepton kinematics, whereas the hadronic tensor, 
\begin{align}
\label{hadronictensor}
W_{\mu\nu}&= \sum_X \,\langle 0 | {J_\mu}^\dagger | X \rangle \,
      \langle X | J_\nu | 0 \rangle \;\delta^{(4)}(p_0 + q - p_X) \ ,
\end{align}
containing all the information on strong interaction dynamics, describes the nuclear response of the target to the transition currents ${J_\mu}$ from the initial state $|0\rangle$, with four momenta $p_0$, to the final state $|X\rangle$, with four momenta $p_X$. 

The nuclear current consists of one- and two-nucleon contributions; the latter arise from processes in which the interaction with the beam particle involves a meson exchanged between the target nucleons and are usually referred to as meson-exchange currents (MEC) 
\begin{equation}
\label{def:curr}
J^\mu = J_1^{\,\mu} + J_2^{\,\mu} +J_3^{\,\mu} = \sum_i j^{\,\mu}_i + \sum_{ j>i} j^{\,\mu}_{ij} +\sum_{k>j>i} j^{\,\mu}_{ijk} \ .
\end{equation}
The relation between the MEC and the nuclear hamiltonian will be discussed in the following section.

\section{Moderate momentum transfer regime}
\label{sec-3}
In the moderate momentum transfer regime, both the initial and final states appearing in  Eq. (\ref{hadronictensor}) 
are eigenstates of the nonrelativistic nuclear Hamiltonian $H$
\begin{equation}
\label{schroedinger}
H |0\rangle=E_0|0\rangle \ \ \ \ , \ \ \ \  H|X\rangle=E_X|X\rangle \ .
\end{equation}
The nuclear Hamiltonian describing the dynamics of the point like nonrelativistic nucleons is given by
\begin{align}
H = \sum_{i=1}^{A} \frac{{\bf p}_i^2}{2m} + \sum_{j>i=1}^{A} v_{ij}
 + \sum_{k>j>i=1}^A V_{ijk} \ .
\label{eq:nucl_h}
\end{align}
In the above equation, ${\bf p}_i$ is the momentum of the $i$-th nucleon, while $v_{ij}$ and $V_{ijk}$ are the two- and three-nucleon potentials, respectively.

For nuclei as large as $^{12}$C, the ground state wave function can be obtained via the Green's Function Monte Carlo (GFMC) approach \cite{kalos:1962,grimm:1971}. It has to be remarked that, within the limits of applicability of NMBT, the GFMC is truly an ab initio approach, allowing {\it exact} calculations of a number of nuclear properties.

In the moderate momentum-transfer regime, the nuclear cross section can be rewritten in terms of the response functions $R_{\mu\nu}(q,\omega)$, obtained from Eq.\eqref{hadronictensor} replacing the components of the current operator with their expressions obtained in the relativistic limit. 

A key feature of the description of neutrino-nucleus interactions at low and moderate momentum transfer is the possibility of employing a set of electroweak charge and current operators consistent with the Hamiltonian of Eq.(\ref{eq:nucl_h}). The nuclear electromagnetic current,   $J^\mu_{\rm em} \equiv (J^0_{\rm em},{\bf J}_{\rm em})$, trivially related to the vector component of the weak current, is constrained by $H$ through the continuity equation \cite{Riska89}
\begin{equation}
\label{continuity}
{\boldsymbol \nabla} \cdot {\bf J}_{\rm em} + i [H,J^0_{\rm em}] = 0 \ .
\end{equation}
Since the two- and three-nucleon potentials $v_{ij}$ and $V_{ijk}$ do not commute with the charge operator $J^0_{\rm em}$, the above equation implies that $J^\mu_{\rm em}$ involves two- and three-nucleon contributions, as shown in Eq.\eqref{def:curr}.

The one-body electroweak operator is obtained from a nonrelativistic expansion of the covariant single-nucleon currents, while the two-body charge and current operators that we employ in our calculations are derived within the conventional meson-exchange formalism \cite{Marcucci:2000, Marcucci:2005}. Nonrelativistic MEC have been used in analyses of a variety of electromagnetic moments and  electroweak transitions of s- and p-shell nuclei at low and intermediate values of energy and momentum transfers \cite{Pervin:2007,Marcucci:2008,Schiavilla:2002,Marcucci:2011,Wiringa:1998,Marcucci:1998,Viviani:2007}, improving on the description of the experimental data with respect to the one-body approximation. 

\begin{figure}
\centering
\includegraphics[width=8cm,clip]{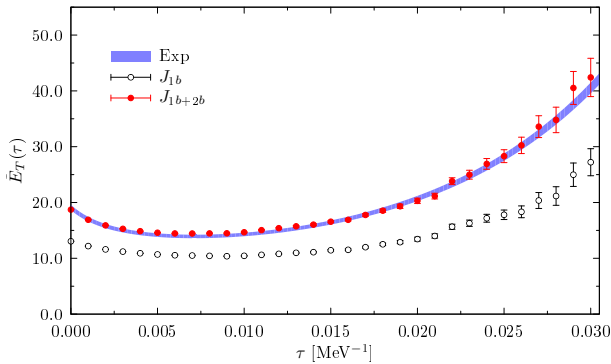}
\caption{Euclidean electromagnetic transverse response function of $^{12}$C at $q = 570$ MeV (adapted from Ref. \cite{Lovato:2015}). The theoretical predictions obtained including only one-body and both one- and two-body transition operators are represented by open circles and solid circles, respectively. The shaded band refers to the Euclidean response extracted from Ref. \cite{Jourdan:1996}.}
\label{fig:12C_euc}      
\end{figure}

We have used the GFMC method to calculate the imaginary-time response function--the so-called Euclidean response function--to the electromagnetic and neutral weak currents of $^{4}$He and $^{12}$C. The Euclidean response function is defined as the Laplace transform of the response,
\begin{equation}
E_{\alpha\beta}(q,\tau) = C_{\alpha\beta}(q)\int_{\omega_{\rm th}}^\infty d\omega\,
 e^{-\tau \omega} R_{\alpha\beta}(q,\omega) \ ,
\label{eq:laplace_def}
\end{equation}
where $\omega_{\rm th}$ is the inelastic threshold and the $C_{\alpha\beta}$ are $q$-dependent normalization factors. In $R_{\alpha\beta}(q,\omega)$ the
$\omega$-dependence enters via the energy-conserving $\delta$-function and the dependence on the four-momentum transfer $Q^2=q^2-\omega^2$ of the
electroweak form factors of the nucleon and $N$-to-$\Delta$ transition in the currents. Once the latter dependence has been removed, as described in Ref. \cite{Lovato:2015}, the Euclidean response can be expressed as a ground-state expectation value,
\begin{equation}
\frac{E_{\alpha\beta}(q,\tau)}{C_{\alpha\beta}(q)}= \frac{\langle 0| O^\dagger_{\alpha}({\bf q}) e^{-(H-E_0)\tau} 
O_{\beta}({\bf q}) |0\rangle}{\langle 0| e^{-(H-E_0)\tau}|0\rangle}\, .
\label{eq:euc_me}
\end{equation}
In the above equation $H$ is the nuclear Hamiltonian of Eq. (\ref{eq:nucl_h}) (here, the AV18+IL7 model), $\tau$ is the imaginary-time, and $E_0$ is a trial energy to control the normalization. The calculation of the above matrix element is carried out with GFMC methods~\cite{Carlson:1992} similar to those used in projecting out the exact ground state of $H$ from a trial state. 

It has to be noted that evaluating the matrix element of Eq. (\ref{eq:euc_me}) for a nucleus as large as $^{12}$C is computationally nontrivial, as it requires computer capabilities and resources available on only forefront computers.

In figure \ref{fig:12C_euc}, taken from Ref. \cite{Lovato:2015}, the electromagnetic transverse Euclidean response function of $^{12}$C, is compared to the one obtained from the analysis of the world data carried out by Jourdan \cite{Jourdan:1996} represented by the shaded band. Two-body MEC contributions substantially increase the one-body electromagnetic transverse response function. This enhancement is effective over the whole imaginary-time region we have considered, with the implication that excess transverse strength is generated by two-body currents not only at energies larger than the one corresponding to the quasi-elastic peak, but also in the quasi-elastic and threshold regions. The full predictions for the transverse Euclidean response functions is in excellent agreement with the experimental data.

The inversion of a Laplace transform subject to statistical Monte Carlo errors, needed to retrieve the energy dependence of the responses, is long known to involve severe difficulties. However, there are techniques developed in condensed matter theory and other contexts, that seem to have successfully overcome the inherent ill-posed nature of the problem. One of these is known as the {\it maximum entropy technique} \cite{Bryan:1990,Jarrell:1996}; we have recently used it to perform stable inversions of the $^4$He electromagnetic Euclidean response.  

\begin{figure}
\centering
\includegraphics[width=8cm,clip]{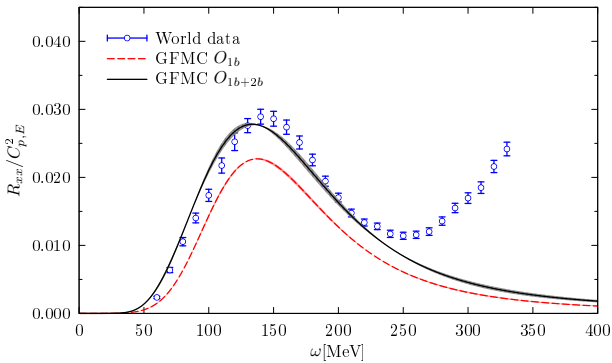}
\caption{Electromagnetic transverse  response functions of $^4$He at $q = 500$ MeV. Experimental data are from Ref. \cite{Carlson:2002}.}
\label{fig:4he_xx}     
\end{figure}

The $^4$He electromagnetic transverse response function (at $q=500$ MeV), obtained from inversion of the corresponding Euclidean response is shown in
figure~\ref{fig:4he_xx}. The inversions are, to a very large degree, insensitive to the choice of default model response~\cite{Lovato:2015} needed by the maximum entropy method. Results obtained with one-body only (dashed line) and (one+two)-body (solid line) currents are compared with an analysis of the experimental world data~\cite{Carlson:2002} (empty circles).  There is excellent agreement between the full theory and experiment. Two-body currents significantly enhance the transverse response function, not only in the dip region, but also in the quasi-elastic peak and threshold regions, providing the missing strength that is needed to reproduce the experimental results.

\section{Relativistic regime}
The dynamical model discussed in section \ref{sec-3} can be also employed to describe the nuclear response in the kinematical region of large momentum transfer, in 
which relativistic effects play a major role. In this regime, the final state includes at least one particle carrying a large
momentum $\sim \mathbf{q}$, and fully relativistic expressions need to be retained for the transition currents. 

The impulse approximation (IA) scheme and the spectral function formalism allow one
to circumvent the difficulties associated with the relativistic treatment of the nuclear final state and current operator,
while at the same time preserving essential features (such as correlations) inherent to the realistic description of nuclear dynamics described in the previous section. The IA scheme is based on the fact that 
in the kinematical region of large momentum transfer, in which $|\mathbf{q}|^{-1} \ll d$, with $d$ being the average nucleon-nucleon separation distance in the target, the nuclear scattering can be reasonably assumed to reduce to the incoherent sum of elementary scattering processes involving individual nucleons.

The ground state of the target nucleus, which does not depend on $\mathbf{q}$, is still an eigenstate of the nuclear Hamiltonian of Eq. \ref{eq:nucl_h}. Within the IA the final state factorizes 
\begin{equation}
|X\rangle = |\mathbf{p}\rangle \otimes |n_{A-1},\mathbf{p}_n\rangle \, .
\label{eq:fs_IA}
\end{equation}
In the above equation, the state $|\mathbf{p}\rangle$ is a plane wave describing a noninteracting nucleon, while $|n_{A-1},\mathbf{p}_n\rangle$ is an eigenstate of the $(A-1)$ nuclear Hamiltonian of Eq. (\ref{eq:nucl_h}) describing the
$(A - 1)$-system recoiling with momentum $\mathbf{p}_n$.
Using the above equation, it is possible to rewrite the nuclear transition matrix element of the one-body current between free nucleon states -- which can be computed exactly, retaining the fully relativistic expressions of the currents -- and the nuclear amplitude involving the target ground state and the state of the spectator $(A-1)$-system. By using Eq. (\ref{eq:fs_IA}) in Eq. (\ref{hadronictensor}), the nuclear cross section can be written in the form \cite{Benhar:2006wy}
\begin{equation}
d\sigma_{IA}=\sum_i\int d^3 k\, dE\, P_i(\mathbf{k},E) d\sigma_i\, ,
\end{equation}
where $d\sigma_i$ is the cross-section describing scattering on the individual i-th nucleon, the momentum and removal energy of which are distributed according to the spectral function $P_i(\mathbf{k},E)$. Highly accurate theoretical calculations of the spectral function can be carried out for uniform nuclear matter, exploiting the simplifications arising from translation invariance \cite{PKE}. The results of these calculations have been combined with the information obtained from coincidence $(e, e^\prime p)$ experiments at moderate energy and momentum transfer, to obtain spectral functions of a variety of nuclei within the local density approximation (LDA) \cite{LDA}.

Note that, because of the strong spatial-spin-isospin correlations that are present among nucleons in the nucleus, the recoiling $(A - 1)$-nucleon system is not necessarily left in a bound, one hole, state. In fact, two hole-one particle states, in which one of the spectator nucleons is excited to the continuum, typically contribute $15-20\%$ of the spectral function normalization, the corresponding strength being located at large momenta and energies. 

Neglecting the contributions of final states involving more than two nucleons in the continuum, the cross section can be written as
\begin{equation}
d\sigma = d\sigma_{1p1h} + d\sigma_{2p2h} \propto L^{\mu\nu}(W_{\mu\nu}^{1p1h}+W_{\mu\nu}^{2p2h})\, .
\end{equation}

Carrying out calculations of nuclear amplitudes combining fully relativistic MEC currents and a description of nuclear dynamics taking into account short range correlations requires a generalization of the factorization ansatz. The authors of Refs. \cite{Benhar:2013oba,Benhar:2015ula} have extended the factorisation ansatz of Eq. (\ref{eq:fs_IA}) to allow for a consistent treatment of the amplitudes involving one- and two-nucleon currents. The resulting expression is
\begin{equation}
|X\rangle = |\mathbf{p}\mathbf{p}^\prime\rangle \otimes | m_{A-2},\mathbf{p}_m\rangle \, ,
\end{equation}
where the states $|\mathbf{p}\mathbf{p}^\prime\rangle$ and $| m_{A-2},\mathbf{p}_m\rangle$ describe two non-interacting nucleons of momenta $\mathbf{p}$ and $\mathbf{p}^\prime$ and the $(A-2)$-particle residual system, respectively. The two-particle two-hole contribution to the hadronic tensor obtained from the extended factorisation ansatz is a sum of three terms,
\begin{equation}
W_{\mu\nu}^{2p2h}=W_{\mu\nu\,11}^{2p2h}+W_{\mu\nu\,12}^{2p2h}+W_{\mu\nu\,22}^{2p2h}\, ,
\end{equation}
where $W_{\mu\nu\,11}^{2p2h}$ and $W_{\mu\nu\,22}^{2p2h}$ involve the squared amplitudes of the matrix elements of one- and two-nucleon currents, while $W_{\mu\nu\,12}^{2p2h}$ describes the interference between the amplitudes involving one- and two- body currents.
\begin{figure}
\centering
\includegraphics[width=8cm,clip]{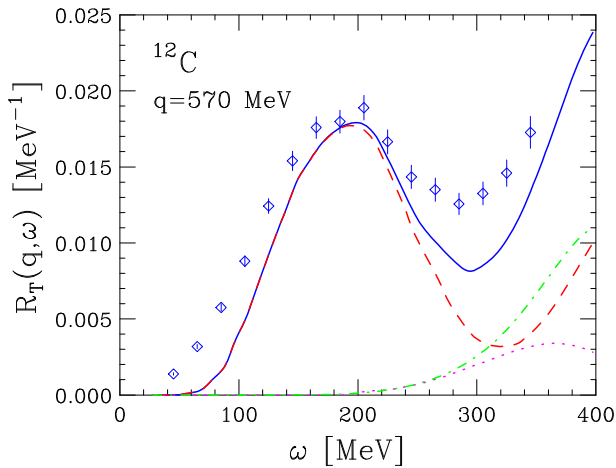}
\caption{Electromagnetic transverse response of $^{12}C$ at momentum transfer $|\mathbf{q}| = 570$ MeV, taken from Ref. \cite{Benhar:2015ula}. The solid line represents the results of the full calculation,
whereas the dashed line has been obtained including only amplitudes involving the one-body current. The contributions arising from the two-nucleon current are illustrated by the dot-dashed and dotted lines, corresponding to the pure two-body current transition probability and the interference term, respectively. The experimental data are taken from Ref. \cite{Jourdan:1996}.}
\label{fig:12c_xx}     
\end{figure}

Figure \ref{fig:12c_xx}, taken from Ref. \cite{Benhar:2015ula}, shows the transverse electromagnetic response of $^{12}C$ at $|\mathbf{q}| = 570$ MeV computed using the carbon spectral function of Ref. \cite{LDA} and approximating the two-hole spectral function of carbon with that of uniform nuclear matter at density corresponding to Fermi momentum $k_F = 221$ MeV. The fully relativistic expression of the MEC described in Refs. \cite{Dekker:1994,DePace:2003} was used, with the same form factors and $\Delta$-width. The solid line represents the results of the full calculation, whereas the dashed line has been obtained including only the amplitudes involving the one-body current. The contributions arising from the MEC are illustrated by the dash-dot and dotted lines, corresponding to the pure two-body current transition probability and the interference term, respectively. The latter turns out to be sizable, its contribution being comparable to the total two-body current response for $\omega < 350$ MeV. It has to be noted that these results still need to be improved, as they do not include the corrections taking into account the effects of final state interactions. The data resulting from the analysis of \cite{Jourdan:1996} are also included for comparison.

A comparison between the results of figure \ref{fig:12c_xx} and the GFMC results of figure \ref{fig:4he_xx} shows distinctive discrepancies in both magnitude and energy dependence of the two-body current contributions. While part of the disagreement is likely to originate from using fully-relativistic and nonrelativistic MEC, as well as from the non relativistic nature of the GFMC calculations, the large interference contribution in the region of the quasi elastic peak observed in figure \ref{fig:4he_xx} may well arise from interference between amplitudes involving the one- and two-body currents and 1p1h final states.

\begin{acknowledgement}
The work of O.B. and N.R. was supported by INFN under grant MANYBODY. The work of AL, SCP, SG, JC and RS a was supported by the U.S. Department of Energy, Office of Science, Office of Nuclear Physics, under contracts DE-AC02- 06CH11357 (AL and SCP), DE-AC52-06NA25396 (SG. and JC), DE-AC05-06OR23177 (RS). The work of AL, SCP, SG and JC was also supported by the NUCLEI SciDAC program. Under an award of computer time provided by the INCITE program, this research used resources of the Argonne Leadership Computing Facility at Argonne National Laboratory, which is supported by the Office of Science of the U.S. Department of Energy under contract DE-AC02-06CH11357.
\end{acknowledgement}

%
 \bibliography{LovatoA_biblio.bib}
%
%
%
%

\end{document}